\documentclass[
 reprint,
 amsmath,amssymb,
 aps,
]{revtex4-2}

\usepackage{graphicx}% Include figure files
\usepackage{dcolumn}% Align table columns on decimal point
\usepackage{bm}% bold math
\usepackage{siunitx} 
\usepackage{comment}
\usepackage[utf8]{inputenc}
\usepackage[T1]{fontenc}

\usepackage{bm}
\usepackage{xcolor}

\usepackage[colorinlistoftodos, textsize=footnotesize]{todonotes}
\usepackage{soul}

\newcommand{\equalcontrib}{\thanks{These authors contributed equally to this work.}}

\begin{document}

\title{Optical Injection and Detection of Long-Lived Interlayer Excitons in van der Waals Heterostructures}

\author{Alperen Tüğen${}^{1}$}
\equalcontrib

\author{Anna M. Seiler${}^{1}$}
\equalcontrib

\author{Arthur Christianen${}^{1}$}

\author{Kenji Watanabe${}^{2}$}

\author{Takashi Taniguchi${}^{3}$}

\author{Martin Kroner${}^{1}$}

\author{Ataç İmamoğlu${}^{1}$}

\affiliation{${}^{1}$Institute for Quantum Electronics, ETH Zurich, CH-8093 Zurich, Switzerland}
\affiliation{${}^{2}$Research Center for Electronic and Optical Materials, National Institute for Materials Science, Tsukuba, Japan}
\affiliation{${}^{3}$Research Center for Materials Nanoarchitectonics, National Institute for Materials Science, Tsukuba, Japan}

\date{\today}

\begin{abstract}

Interlayer excitons in semiconducting bilayers separated by insulating hBN layers constitute a promising platform for investigation of strongly correlated bosonic phases. Here, we report an optical method for the generation and characterization of long-lived interlayer excitons. We confirm the presence of tightly bound interlayer excitons by measuring 1s and 2s intralayer excitons in each layer concurrently. Using a pump-probe technique, we find interlayer exciton lifetimes up to $\qty{8.8}{\micro\second}$, increasing with the thickness of the hBN. With optical access to long-lived interlayer excitons, our approach provides a new route to explore degenerate Bose--Fermi mixtures of excitons and itinerant electrons with high spatial and temporal resolution.

\end{abstract}

\maketitle

Transition metal dichalcogenide (TMD) heterobilayers have recently emerged as a promising platform for creating long-lived interlayer excitons, bound electron-hole pairs, whose constituents reside in different layers. With their large binding energy and a built‑in out‑of‑plane dipole, they offer a route to investigate many-body phenomena ranging from superfluidity of dipolar excitons \cite{lu2011strongly,aikawa2012bose,alloing2014evidence,combescot2017bose} to Bose–Fermi mixtures exhibiting exotic electron pairing mechanisms \cite{zerba2024realizing,von2024superconductivity} in van der Waals heterostructures. However, accessing these collective states depends on controlling the lifetime, density, and interactions of interlayer excitons \cite{high2012spontaneous,alloing2014evidence}. 

Early demonstrations of long-lived interlayer excitons relied on the electrical injection of charge carriers into separate TMD layers \cite{ma_strongly_2021,qi_thermodynamic_2023,nguyen_perfect_2025,qi_perfect_2025}. Capacitance spectroscopy and Coulomb‑drag measurements provided compelling evidence for interlayer exciton formation. 
However, these measurements offer no information on time dynamics and spatial distributions. In parallel, optical pumping in aligned MoSe\textsubscript{2}–WSe\textsubscript{2} bilayers, either in direct contact or separated by one or two hexagonal boron nitride (hBN) layers \cite{rivera2015observation, jauregui2019electrical, shimazaki_strongly_2020, mahdikhanysarvejahany2022localized} has been explored. 
Photoluminescence (PL) from such devices has revealed lifetimes up to $\qty{1.9}{\micro\second}$ \cite{cutshall2025imaging}. However, achieving longer lifetimes and higher densities requires thicker hBN spacer layers to suppress residual interlayer tunneling and radiative recombination \cite{snoke2011coherence}. This, in turn, significantly reduces PL emission, rendering direct optical measurements of the interlayer exciton dynamics untenable.

Here, we demonstrate an optical scheme that overcomes the limitations of both electrical and PL-based injection and detection schemes. While inserting hBN spacer with up to seven layers to suppress interlayer tunneling, we inject high densities of electrons and holes into separate MoSe\textsubscript{2} and WSe\textsubscript{2} layers via nonresonant optical pumping, without requiring layer-selective contacts or complex gating geometries \cite{ma_strongly_2021, qi_thermodynamic_2023}. Using time-resolved reflection spectroscopy and the 2s exciton resonance as a spectroscopic probe, we observe interlayer exciton formation and relaxation dynamics in samples with negligible interlayer exciton PL. Under these conditions, we achieve interlayer exciton densities up to $n_\text{IX} \approx \qty{1e12}{\per\square\centi\metre}$  and lifetimes of $\qty{8.8}{\micro\second}$. This combination of high density, long lifetime, and optical control opens the door to studying strongly interacting interlayer exciton gases and exploring collective excitonic phenomena.

\begin{figure*}
\includegraphics{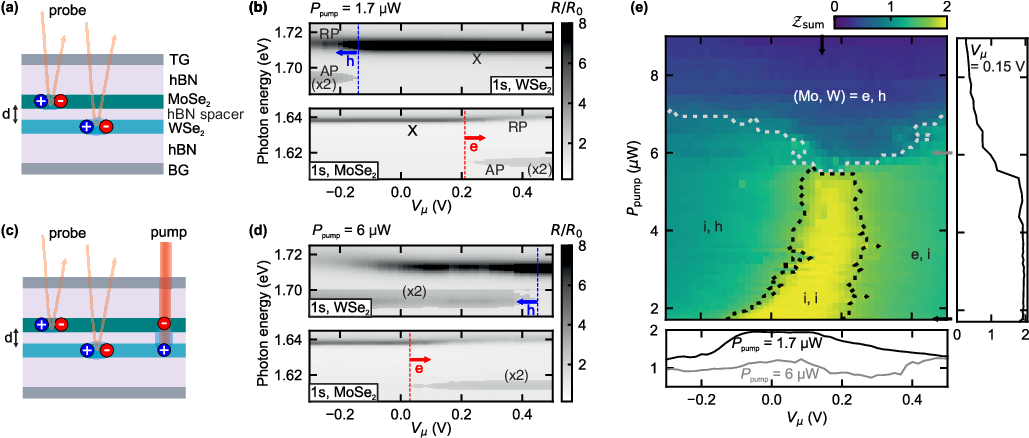}
\caption{\label{fig:1}
(a) Schematic of the device structure.
(b) Normalized reflection spectra ($R/R_0$) of the MoSe\textsubscript{2} and WSe\textsubscript{2} layers as a function of gate voltage $V_\mu$. Resonances corresponding to the repulsive polaron (RP), attractive polaron (AP), and neutral exciton (X) are indicated. Blue and red dashed lines mark the onset of hole and electron doping, respectively. A weak pump power was present during acquisition (\(P_{\rm pump}\) =  $\qty{1.7}{\micro\watt}$); its effect on the spectra is negligible.
(c) Schematic under pump--conditions.
(d) $R/R_0$ at finite pump power (\(P_{\rm pump}\) =  $\qty{6}{\micro\watt}$). 
(e) The sum of the peak 1s X (RP) reflection contrasts of the MoSe\textsubscript{2} and WSe\textsubscript{2} layers, each normalized to its respective reflection contrast in the absence of charges, $\mathcal{Z}_{\mathrm{sum}}$, is shown as a function of $V_\mu$ and $P_\mathrm{pump}$. 
Distinct charge configurations are labeled as (i, i), (i, h), (e, i), and (h, e), representing combinations of intrinsic (i), hole (h), and electron (e) doping in the Mo and W layers, respectively. 
The dotted black contour marks the region where the normalized X (RP) reflection contrast in both layers exceeds 0.7, indicating approximate charge neutrality; note, however, that small but finite doping persists near the boundary of the (i, i) region. The dotted light grey contour outlines the region with opposite doping in the two layers, where the normalized X (RP) reflection contrast in each layer falls below 0.7. Arrows indicate the positions of vertical and horizontal linecuts. Data in panels (b)--(e) were acquired using Device 1.}
\end{figure*}

We investigate three TMD bilayer devices with varying hBN spacers (see Fig.~\ref{fig:1}a). In the main text, we present data from Devices~1 and~2, both MoSe\(_2\)–WSe\(_2\) bilayers with hBN spacers of 1- and 3--5-layers, while results from Device~3 (MoS\(_2\)–WSe\(_2\), 5--7 layer hBN spacer), are provided in the Supplemental Information (SI). In contrast to previous optical studies, all TMD layers within a device are angle-misaligned, suppressing interlayer hybridization, and resulting in negligible interlayer PL~\footnote{Device 1 shows very weak interlayer PL, whereas no interlayer PL is observed in Devices 2 and 3; see Fig. S2}. To determine the charge configuration in the individual layers, we perform reflection spectroscopy. 
The upper panel of Fig.~\ref{fig:1}b shows spectra of the MoSe$_2$ and WSe$_2$ 1s exciton resonances as a function of the gate voltage $V_\mu$ (controls the chemical doping of the layers, see SI for details), measured in Device~1. 
Upon injection of itinerant charge carriers, the 1s exciton evolves into two distinct resonances (Fig.~\ref{fig:1}b): the repulsive polaron (RP) and the attractive polaron (AP) \cite{sidler_fermi_2017}. The left (right) side of the blue (red) dashed line in Fig. \ref{fig:1}b corresponds to hole (electron) injection into the WSe$_2$ (MoSe$_2$) layer. The voltage range between the dashed lines marks the charge-neutral regime, where neither layer is doped.

Unlike previous experiments that bridged this energy gap using large electric fields and an interlayer bias ~\cite{ma_strongly_2021, qi_thermodynamic_2023, qi_perfect_2025, nguyen_perfect_2025}, we employ nonresonant optical pumping to inject charge carriers into the TMD bilayer. We illuminate the sample with a continuous-wave laser at 635 nm, above the bandgap (Fig.~\ref{fig:1}c) \textcolor{black}{with pump and probe focused onto the same spot}. Remarkably, as we show below, this optical doping technique enables simultaneous electron- and hole-doping in opposite layers even for devices with thicker hBN spacers, overcoming the intrinsic band-offset limitations without the need for applying an electrical bias. 
Fig.~\ref{fig:1}d displays the reflection spectra under continuous pumping at laser power $P_{\text{pump}} = \qty{6}{\micro\watt}$. Here, the charge-neutral region vanishes, and we observe simultaneous electron doping in the MoSe$_2$ layer and hole doping in the WSe$_2$ layer. This is indicated by the reversal in the positions of the red and blue dashed lines~\footnote{We emphasize that the two types of doping can be spectrally distinguished in WSe$_2$ \cite{wang2020observation}: the presence of a single attractive polaron (AP) resonance in the WSe$_2$ spectrum unequivocally indicates hole doping, since electron doping of WSe$_2$ leads to two AP features associated with the singlet and triplet trions. In contrast, MoSe$_2$ displays a single AP peak under both electron and hole doping. Additional confirmation comes from Device 3, a MoS$_2$/WSe$_2$ heterostructure, where the MoS$_2$ layer clearly exhibits two AP features under electron doping, consistent with our interpretation of the spectra in Devices 1 and 2 (see Fig. S6).}.

\begin{figure*}
\includegraphics{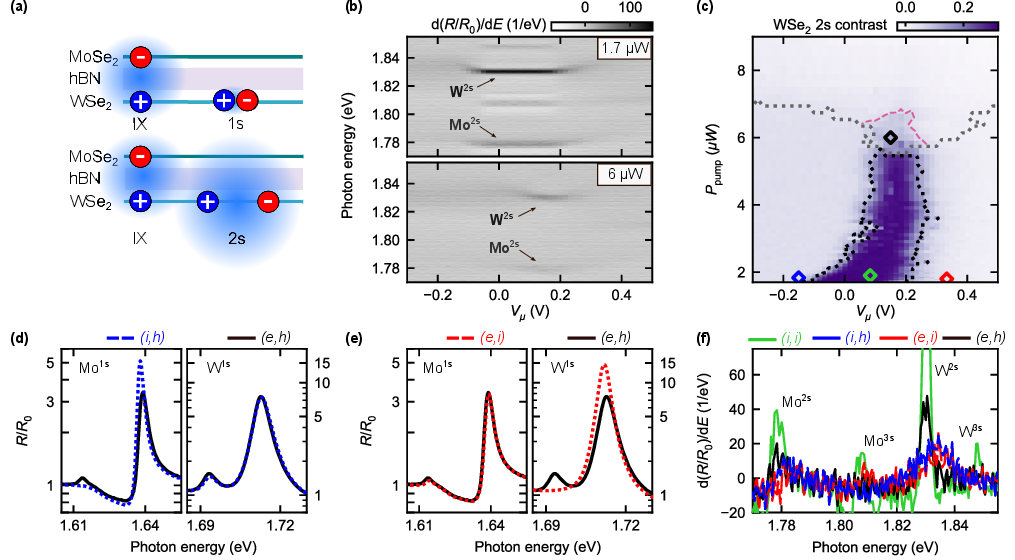}
\caption{\label{fig:2} (a) Schematic of the 1s, 2s intralayer excitons and interlayer exciton (IX). 
(b) Differential reflection spectra ($\mathrm{d} (R/R_0) / \mathrm{d} E $) plotted as a function of gate voltage \(V_\mu\), in the energy range of the Rydberg excitons. Data are shown for pump powers of $P_{\text{pump}} = \qty{1.7}{\micro\watt}$ (upper panel) and $P_{\text{pump}} = \qty{6}{\micro\watt}$ (lower panel). MoSe$_2$ and WSe$_2$ 2s Rydberg excitons are labeled (see Fig. S7 for discussion on 3s Rydberg excitons).
(c) Colormap of WSe\textsubscript{2} 2s contrast as a function of $V_\mu$ and $P_\mathrm{pump}$. The dashed black and grey contour lines delineate different charge configurations, defined as in Fig.~\ref{fig:1}e. The dashed pink contour marks the boundary where the WSe\textsubscript{2} 2s contrast falls below 0.055 ($20\%$) in the region where the two layers are oppositely doped.
(d,e) Log-scale spectral linecuts at distinct doping configurations (see diamond markers in panel~(c)). The blue and red curves correspond to only hole doping in WSe\textsubscript{2} and only electron doping in MoSe\textsubscript{2}, respectively, with the opposite layer remaining charge neutral. The black curve shows simultaneous doping of both layers, with individual charge carrier densities matching those in the red and blue cases.
(f) Differential reflection spectra ($\mathrm{d} (R/R_0) / \mathrm{d} E$) plotted across the Rydberg exciton resonances. The green curve, measured at charge neutrality (green diamond), shows the bare Rydberg excitons. The 2s resonance disappears when only one layer is doped (red and blue curves), but remains visible without a significant spectral shift when both layers are doped simultaneously (black curve). Data in panels (b)--(f) were acquired using Device 1.}
\end{figure*}

To visualize the overall charge configuration, Fig.~\ref{fig:1}e shows the sum of the peak 1s X (RP) reflection contrast of MoSe\textsubscript{2} and WSe\textsubscript{2} layers normalized to the individual reflection contrast of each layer in the absence of charges ($\mathcal{Z}_{\mathrm{sum}}$) as a function of $V_\mu$ and $P_{\text{pump}}$ (see SI for details). The yellow region labeled as (i,i) and outlined by a black-dotted contour denotes the approximate charge-neutral regime, where both excitons reach maximum contrast. At higher $P_\mathrm{pump}$, a dark blue region appears above a light gray dotted contour line, indicating dual-layer doping with opposite charge carriers (e,h).
In the surrounding regions, either WSe\(_2\) is hole doped (i,h) or MoSe\(_2\) is electron doped (e,i), with the opposite layer remaining charge neutral. \textcolor{black}{Details on the asymmetry between hole and electron doping are provided in the SI.}

\begin{figure}[b]
\includegraphics{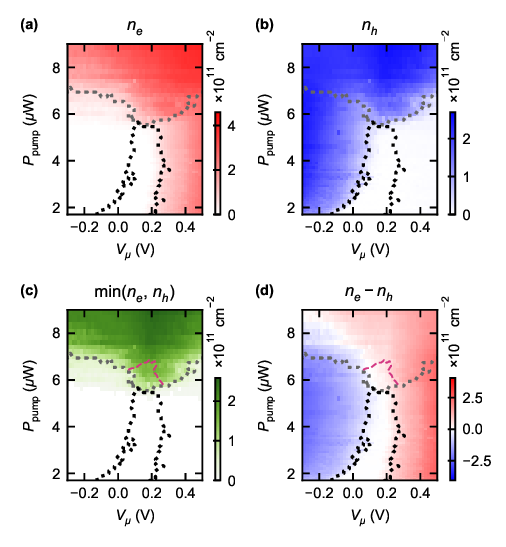}% Here is how to import EPS art
\caption{\label{fig:fig_3} Charge carrier densities extracted from the optical spectra measured in Device 1 as a function of the gate voltage, \(V_\mu\), and pump power, \(P_{\rm pump}\): (a) electron density, \(n_e\), in the MoSe\(_2\) layer; (b) hole density, \(n_h\), in the WSe\(_2\) layer; (c) minimum of \(n_e\) and \(n_h\) (min$(n_e, n_h)$); (d) excess charge \(n_e - n_h\). Dotted contour lines denote the same boundaries depicted in Figs. \ref{fig:1}e and \ref{fig:2}c. 
}
\end{figure}

\begin{figure}[b]
\includegraphics{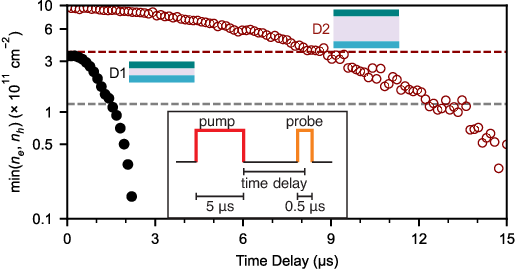}% Here is how to import EPS art
\caption{\label{fig:fig_4} Time evolution of the minimum of the electron and hole densities, \(\min(n_e,n_h)\), as a function of time delay between the pump and probe pulses (see inset). Data points from Device 1 (D1, monolayer hBN spacer) are shown as black filled circles, while those from Device 2 (D2, 3–5‑layer hBN spacer) are shown as dark red open circles. The dashed lines mark the density at which \(\min(n_e,n_h)\) has decayed to \(1/e\) of its initial value; the corresponding delay times are $\qty{1.4}{\micro\second}$ for Device 1 and $\qty{8.8}{\micro\second}$ for Device 2.}
\end{figure}

%Having established the charging configuration in both layers, 
Next, we investigate whether the injected electrons and holes form bound interlayer exciton states, or remain unbound. Prior theoretical work~\cite{amelio_polaron_2023,qi_thermodynamic_2023} demonstrated that the modification of the 1s exciton-polaron resonances can be used to differentiate between bound and unbound electron-hole pairs; however, the associated spectral shifts are rather small and can be masked by exciton line broadening. The origin of this relatively weak dependence is the small Bohr radius of 1s intralayer excitons, rendering them insensitive to the charges in the neighboring TMD layer (see the sketch in the upper panel of Fig.~\ref{fig:2}a). In contrast, Rydberg exciton states, such as the 2s state, exhibit a significantly larger Bohr radius (lower panel of Fig.~\ref{fig:2}a), making them more sensitive to the surrounding electrostatic environment~\cite{xu_correlated_2020}. As such, the 2s exciton reflection contrast can be expected to respond differently depending on whether the charge carriers are (tightly) bound as interlayer excitons or remain free unbound particles.

Spectrally, we can resolve MoSe$_2$ and WSe$_2$ 2s and 3s excitons in the charge-neutral regime (Fig.~\ref{fig:2}b). In the case of free carriers residing in the same layer, 2s excitons exhibit a blue shift and loss of oscillator strength even for small carrier densities. 
\textcolor{black}{When charges are located in the opposite layer, the 2s-resonance still quickly disappears, but part of its oscillator strength is transferred to a new red-shifted resonance which arises from the hybridization with the 2p-states \cite{xu_correlated_2020,popert_optical_2022,kim2025excitons,christianen2025hybridzation}}. 

Interestingly, we observe that in the regime where both layers are doped under finite $P_\mathrm{pump}$, the MoSe$_2$ and WSe$_2$ 2s exciton features persist with reduced intensity \textcolor{black}{and only a small spectral redshift (see SI)}. To quantify this behavior across the charge map, we extract the WSe$_2$ 2s contrast, defined as the peak-to-peak amplitude of the dispersive Lorentzian lineshape associated with the WSe$_2$ 2s exciton (Fig.~\ref{fig:2}c).

Figures~\ref{fig:2}d–f show differentiated spectral linecuts taken at different positions in the phase diagram. In the (e,h) regime (black diamond), the 2s exciton contrast remains finite, even though the charge densities match those in the red and blue diamond regions, where the 2s resonance vanishes. Moreover, the 2s resonance energy in the (e,h) region coincides with that at charge neutrality (green diamond). This robustness of the 2s resonance in the presence of dual-layer doping stands in stark contrast to its drastic modification due to free charges in either layer, and points to a distinct phase: \textcolor{black}{the formation of interlayer excitons. Because interlayer excitons are charge neutral and do not generate strong in-plane electric fields, they do not directly hybridize the 2s and 2p-states. Instead, the leading-order interaction between the 2s and interlayer excitons is a weak attractive van der Waals interaction (see SI), leading to a small density-dependent redshift}.

\textcolor{black}{We note that the persistence of the 2s was consistently observed at different spots within a device as well as across multiple devices, indicating that spatial inhomogeneities are unlikely to account for the observed feature. Since the measured electron and hole trion binding energies in the two layers agree well with previously studied monolayer devices, we can also rule out strong trapping of the optically introduced charges~\cite{kiper2025confined} as an explanation for the robustness of 2s excitons.}

We further observe that the 2s exciton contrast diminishes progressively with increasing $P_\mathrm{pump}$. \textcolor{black}{This behavior is expected: as the interlayer exciton spacing decreases and approaches the 2s Bohr radius, }\textcolor{black}{the 2s state will cease to be bound.} The fact that the 3s exciton resonance vanishes at even lower interlayer exciton densities supports our explanation (Fig. S7). \textcolor{black}{We emphasize that the 2s signal quenches at densities far below the expected interlayer exciton Mott density of $\approx 4 \times 10^{12}~\text{cm}^{-2}$ \cite{amelio_polaron_2023, ma_strongly_2021}}.

Figures~\ref{fig:fig_3}a and \ref{fig:fig_3}b show maps of the electron and hole densities, \(n_e\) and \(n_h\), as functions of \(V_\mu\) and \(P_\mathrm{pump}\) (see SI). From these data we extract the minimum of $(n_e, n_h)$ (\(\min(n_e, n_h)\)) and the excess charge density \(n_e - n_h\), plotted in Figures~\ref{fig:fig_3}c and \ref{fig:fig_3}d.  

In the region where the 2s contrast is visible and the excess charge approaches zero (pink dashed contour), all carriers are paired and thus form interlayer excitons. Thus, the interlayer exciton density is equal to \(\min(n_e, n_h)\). Outside the pink contour, interlayer excitons may still be present alongside free carriers, but the 2s spectroscopic signature lacks sufficient contrast to confirm their presence. Here, \(\min(n_e, n_h)\) provides an upper bound on the attainable interlayer exciton density. In Device~1, where the MoSe$_2$ and WSe$_2$ layers are separated by a monolayer of hBN, the min$(n_e, n_h)$ reaches up to $\qty{3e11}{\per\square\centi\metre}$. By contrast, in Device~2, which features an hBN spacer thickness of 3–5 layers, min$(n_e, n_h)$ 
reaches $\qty{1e12}{\per\square\centi\metre}$ (Fig. S10).

Our optical pumping scheme provides access to interlayer exciton dynamics. We probe these dynamics using time-resolved reflection spectroscopy, with synchronized modulation of the pump and probe beams that allows for programmable delays between pulses (see SI for details). In Fig.~\ref{fig:fig_4}, we present the evolution of $\min(n_e, n_h)$ on a logarithmic scale as a function of pump--probe delay, enabling us to quantify the lifetime of the interlayer exciton state at a representative point in the phase diagram, where $\min(n_e, n_h)$ is finite, the total excess charge is negligible, and the WSe$_2$ 2s exciton is present. The decay of the $\min(n_e, n_h)$ signal with increasing time delay reflects the relaxation of interlayer excitons after the pump laser is turned off. Although the decay is clearly nonexponential, we extract a 1/e lifetime of $\qty{1.4}{\micro\second}$ for Device 1, which contains a monolayer hBN spacer. In contrast, a significantly longer lifetime of $\qty{8.8}{\micro\second}$ is observed for Device~2, which has a thicker hBN spacer of 3--5 layers. 

This prolonged lifetime is consistent with expectations: increased spacer thickness suppresses interlayer tunneling and reduces the radiative decay rate, thereby extending the interlayer exciton lifetime~\cite{cutshall2025imaging}. Surprisingly, the decay becomes faster at lower carrier densities~\footnote{The parabolic trend in the logarithmic plot in Fig.~\ref{fig:fig_4} is clearly visible; similar nonexponential time dependence has also been observed by Cutshall et al. \cite{cutshall2025imaging} in a time-resolved PL measurement}. We speculate that the shortened lifetime may stem from interlayer excitons becoming more tightly bound as their density is lowered: Since the radiative decay rate scales inversely with the square of the Bohr radius, we would expect the lifetime to become shorter. Lastly, we emphasize that our measurements cannot rule out a dominant, sample-dependent, nonradiative decay mechanism.

\textcolor{black}{The generation of interlayer excitons within a focused optical spot, combined with the ability to probe their dynamics on picosecond timescales~\cite{uto2024interaction}, provides a powerful platform for exploring degenerate dipolar excitons with high spatial and temporal resolution. Crucially, it allows spatially controlled exciton generation without the need for external electric fields, which are typically required in schemes requiring bias voltage for interlayer exciton generation~\cite{ma_strongly_2021, qi_thermodynamic_2023}. This capability opens the door to creating multiple excitonic reservoirs within a single device, paving the way for Josephson-like experiments~\cite{khorana1969ac, lagoudakis2010coherent, abbarchi2013macroscopic}. In particular, optically inducing high-density exciton populations in two gate-defined traps ~\cite{rapaport2005electrostatic,gartner2007micropatterned} connected by a narrow channel could provide a direct route to optically probing \cite{regan2020mott} coherent tunneling and phase dynamics between condensate regions.}

\begin{acknowledgments}
We thank Xiaobo Lu for fabricating Device 1. We thank Ivan Amelio, Haydn S. Adlong and Igor Khanonkin for inspiring discussions. This work was supported by the Swiss National Science Foundation (SNSF) under Grant No. 200021-204076. A.M.S. and A.C. acknowledge funding from an ETH Postdoctoral Fellowship. K.W. and T.T. acknowledge support from the JSPS KAKENHI (Grant Numbers 21H05233 and 23H02052), the CREST (JPMJCR24A5), JST and World Premier International Research Center Initiative (WPI), MEXT, Japan.
\end{acknowledgments}

%\bibliography{Citations}
%\bibliographystyle{apsrev4-2}
%

\end{document}

\title{Supplementary material:\\
Optical Injection and Detection of Long-Lived Interlayer Excitons in van der Waals Heterostructures}

%\keywords{Suggested keywords}%Use showkeys class option if keyword

                          %display desired
\maketitle

%\tableofcontents

\section{Device Fabrication and Characterization}
\label{sec:device_fab}

Monolayer flakes of MoSe$_2$, WSe$_2$, and MoS$_2$, as well as few-layer graphene and hexagonal boron nitride (hBN), were mechanically exfoliated from bulk crystals onto silicon (Si) substrates with a 285~nm thermally grown silicon dioxide (SiO$_2$) capping layer. Flakes were identified and characterized using bright-field optical microscopy, with their thicknesses estimated based on optical contrast. The thickness of the hBN spacer in Device~1 was later confirmed by photoluminescence (PL) measurements (Fig. \ref{fig:PL}). The heterostructures were assembled using a standard dry transfer technique involving a poly(bisphenol A carbonate) (PC)-coated hemispherical polydimethylsiloxane (PDMS) stamp. Individual flakes were sequentially picked up and stacked, and then transferred onto a Si/SiO$_2$ (285~nm) substrate.

\begin{figure*}
\includegraphics{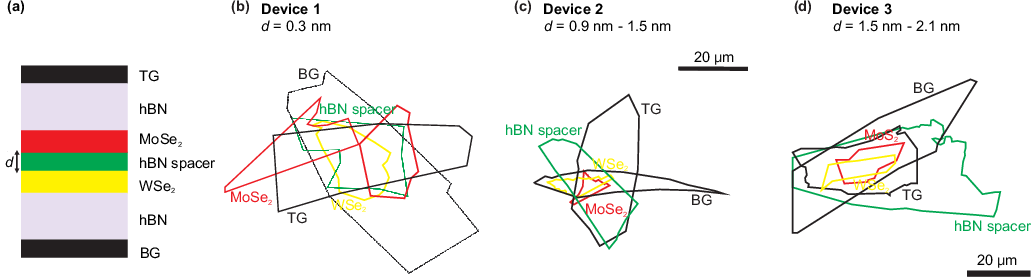}
\caption{\label{fig:devices}
(a) Schematic of the device structure. (b–d) Optical microscope images of the three devices. Top gates (TG) and bottom gates (BG) are outlined in black; MoSe\textsubscript{2}/ MoS\textsubscript{2} and WSe\textsubscript{2} flakes are outlined in red and yellow, respectively; and the hBN spacers are outlined in green. Device 1 (b) contains a single-layer hBN spacer (0.3~nm thick), Device 2 (c) has a 3–5 layer spacer (0.9–1.5~nm), and Device 3 (d) has a 5–7 layer spacer (1.5–2.1~nm). Note that the MoSe\textsubscript{2} flake in Device~1 contains two monolayer regions, which are connected by a thicker region of bulk MoSe\textsubscript{2} (not outlined in the image).
}
\end{figure*}

\begin{figure*}
\includegraphics{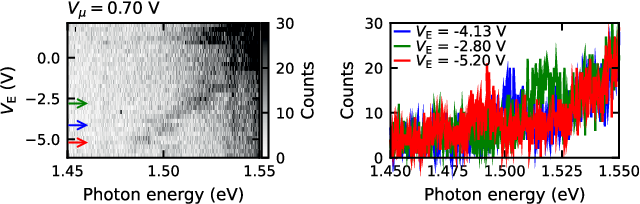}
\caption{\label{fig:PL}
Photoluminescence (PL) intensity map measured on Device 1 as a function of photon energy and applied electric field \( V_E \), at fixed charge carrier density \( V_{\mu} = 0.70\,\mathrm{V} \).
Colored arrows indicate selected values of \( V_E \), for which the corresponding PL spectra are shown in the right panel.
The integration time was 90 seconds. From the slope of the emission energy as a function of \( V_E \), we extract an electric dipole length of 0.88\,nm, consistent with a single monolayer of hBN as the spacer.
}
\end{figure*}

A total of three heterostructures were fabricated. Devices~1 and~2 are based on WSe$_2$/MoSe$_2$ heterobilayers, while Device~3 consists of a WSe$_2$ and a MoS$_2$ flake. The two TMD layers in each device were separated by a thin hBN spacer layer with thicknesses ranging from 0.3 to 2.1~nm. Specifically, Device~1 (Fig.~\ref{fig:devices}b) contains a single-layer hBN spacer (approximately 0.3~nm); Device~2 (Fig.~\ref{fig:devices}c) has a 3–5 layer spacer (approximately 0.9–1.5~nm); and Device~3 (Fig.~\ref{fig:devices}d) contains a 5–7 layer spacer (1.5–2.1~nm). Each device further includes hBN layers as top and bottom gate dielectrics, with thicknesses as follows: Device~1 — top hBN: 40~nm, bottom hBN: 50~nm; Device~2 — top: 17~nm, bottom: 25~nm; Device~3 — top: 25~nm, bottom: 25~nm. Few-layer graphene was used as for top and bottom gate electrodes in all devices. In addition, Devices~1 and~3 include graphite flakes that serve as electrical contacts to the TMD layers.

Metallic contacts to the gate electrodes and to the TMD or graphite contact layers were defined using optical lithography, followed by electron beam evaporation. A 5~nm titanium (Ti) adhesion layer and an 85~nm gold (Au) layer were deposited to form electrical contacts to the graphite gates and contact electrodes. In Device~2, the MoSe$_2$ layer was contacted using a 30~nm bismuth (Bi) layer capped with 20~nm of Au, while the WSe$_2$ flake was contacted using 15~nm of platinum (Pt).

Optical microscopy images of the fabricated devices are shown in Fig.~\ref{fig:devices}.

\section{Optical Spectroscopy}
\label{sec:device_fab}

All reflection measurements were conducted at cryogenic temperatures of approximately $T \approx 4~\mathrm{K}$ using a variety of confocal microscope setups.

Most measurements on Device 1, along with all measurements on Device 2, were performed in a dry cryostat (attodry800) with free-space optical access. For reflection spectroscopy, we utilized a supercontinuum laser spanning a wavelength range of 655–740~nm, in combination with a single-mode, fiber-coupled broadband light-emitting diode (LED) centered at 760~nm with a 3dB bandwidth of 20~nm. The light sources were focused onto the sample using a high-numerical-aperture microscope objective (LT-APO/VISIR/0.82, $NA \approx 0.8$). The typical incident optical power on the sample was approximately 70~nW. Reflected light was collected through the same objective, separated from the excitation path by a beamsplitter, and directed to a spectrometer for detection, with exposure times ranging from 100~ms to 1~s. To optically generate charge carriers in the system, an additional pump laser at 635~nm operating in the microwatt power range was confocal with the broadband probe. The excitation power was finely controlled using a fiber-coupled electronic variable optical attenuator (VOA).

Measurements on Device~3 were performed in a helium bath cryostat, using the same optical setup as described above, but with a different microscope objective (LT-APO\slash LWD\slash VISIR\slash 0.65, \mbox{$NA \approx 0.65$}).

Landau level spectroscopy measurements, as shown in Fig. \ref{fig:ll_calibration}, were carried out in a helium bath cryostat equipped with a superconducting magnet capable of applying magnetic fields up to 14~T perpendicular to the sample surface. The optical setup was identical to that used in the dry cryostat, with the addition of polarization optics. Specifically, linear polarizers, half- and quarter-wave plates were placed in both the excitation and detection paths. This configuration enabled control over the polarization of the incident beam and selective detection of circularly polarized components in the reflected signal.

For the time-dependent reflection measurements, the pump laser was modulated into a square-wave pulse train using an external function generator operating at a frequency of 100~kHz with a duty cycle of 5\%. To enable synchronized probing, an acousto-optic modulator (AOM) in a double-pass geometry was integrated into the probe path and driven at the same modulation frequency \cite{reinhard2013strong}. By precisely controlling the relative delay between the pump and probe pulses, we were able to selectively measure the system’s optical response at defined time intervals following excitation.

\section{Electrical Control of the Devices}
\label{sec:electric_control}

Our devices are equipped with graphite top and bottom gates, which are electrically connected to a Keysight source-measure unit. The chemical potential and the perpendicular electric field across the heterostructure were tuned independently via these gates.

To control the system, we define the following combinations of gate voltages:

\begin{align}
V_\mu &= \alpha_\mu V_\mathrm{bg} + (1 - \alpha_\mu) V_\mathrm{tg}, \\
V_E   &= \frac{1}{2} (V_\mathrm{bg} - V_\mathrm{tg}),
\end{align}

where $V_\mathrm{bg}$ and $V_\mathrm{tg}$ are the bottom and top gate voltages, respectively, and $\alpha_\mu = d_\mathrm{tg} / (d_\mathrm{bg} + d_\mathrm{tg})$ accounts for the relative thicknesses of the hBN layers, with $d_\mathrm{bg}$ and $d_\mathrm{tg}$ denoting the bottom and top hBN thicknesses.

For Device~1, we confirmed $\alpha_\mu = 0.37$ based on the slope of constant-doping contours in a two-dimensional map of optical spectra measured as a function of $V_\mathrm{tg}$ and $V_\mathrm{bg}$. The conversion from $V_\mu$ to charge carrier density $n$ is discussed in Section \ref{sec:carrier_density}.

\section{Analysis of the Reflection Spectra}
\label{sec:analysis_opt_spectra}

We extract the reflection contrast by normalizing the measured reflection spectra to a reference spectrum that captures the background contribution from the broadband light source. The reference spectrum, $R_0(E)$, is acquired at a fixed spatial position on the sample, but under gate voltage conditions that suppress sharp optical resonances. Specifically, we apply large gate voltages to drive the system far from excitonic resonance conditions and into a high-doping regime, where exciton and polaron features are significantly broadened as a result of dynamical screening. In this regime, the reflection is primarily governed by the spectral profile of the incident light, making $R_0(E)$ a reliable baseline for normalization.

The reflectance contrast is then calculated as:
\[
R_\mathrm{c}(E) = \frac{R(E)}{R_0(E)},
\]
where $R(E)$ is the reflectance spectrum measured at a given gate voltage and $E$ denotes the photon energy.

\begin{figure*}[!ht]
\includegraphics{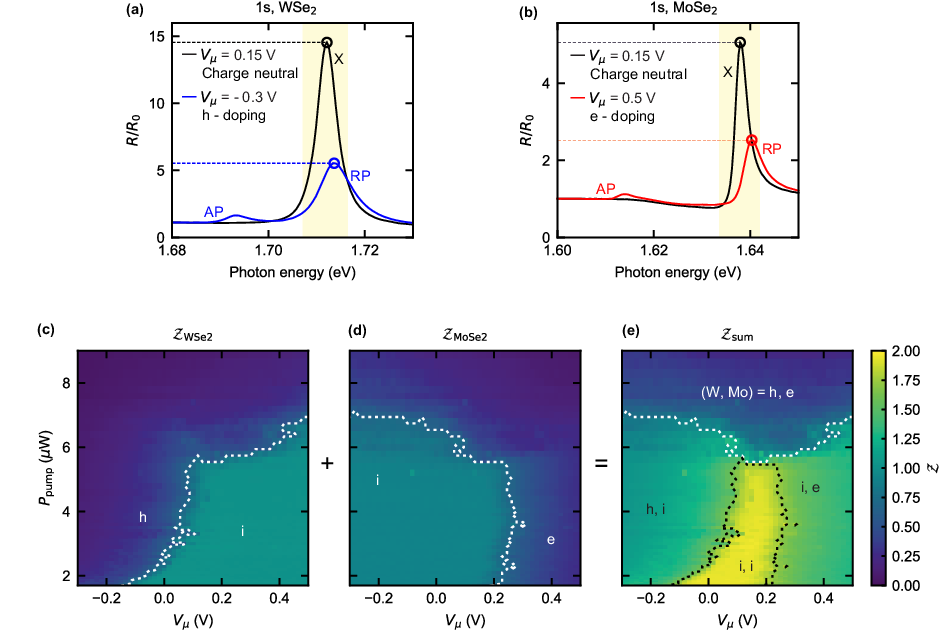}
\caption{\label{fig:doping}
(a,b) Differential reflectivity spectra ($R/R_0$) of the MoSe\textsubscript{2} (a) and WSe\textsubscript{2} (b) layers as a function of photon energy, measured under charge-neutral and doped conditions at the lowest pump power in Device~1. Distinct neutral exciton (X), repulsive polaron (RP), and attractive polaron (AP) resonances are labeled. Electron (e) doping is applied to MoSe\textsubscript{2} and hole (h) doping to WSe\textsubscript{2} via the gate voltage $V_\mu$. The X/ RP peaks are marked.  
(c,d) Pump power ($P_\mathrm{pump}$) vs.\ gate voltage ($V_\mu$) maps of the peak 1s X (RP) reflection contrast of Moe\textsubscript{2} (c) and WSe\textsubscript{2} (d) layers normalized to the individual reflectance contrast of each layer in the absence of charges ($\mathcal{Z}$). Regions dominated by intrinsic (i), electron (e), or hole (h) doping are labeled and separated by black dashed contours where $\mathcal{Z}$ drops below 0.7.  
(e) Sum of the two maps illustrates the full charge configuration of the heterostructure. Regions are labeled according to the doping in each layer. The white dotted contour marks the charge-neutral region (i, i), where $\mathcal{Z}$ in both layers falls below 0.7. The black dotted contour highlights regions of opposite doping, where $\mathcal{Z}$ in both layers exceeds 0.7.
}
\end{figure*}

We then used these normalized reflection spectra, $R/R_0$, to determine the charge configuration in our devices as a function of the gate voltage $V_\mu$. Figures~\ref{fig:doping}a,b show normalized reflection spectra of the MoSe\textsubscript{2} and WSe\textsubscript{2} layers in Device~1. In the charge-neutral regime, both layers display sharp exciton resonances, which correspond to tightly bound electron–hole pairs in the absence of free carriers. As $V_\mu$ is tuned to introduce electron doping in MoSe\textsubscript{2} or hole doping in WSe\textsubscript{2}, these excitonic features evolve into two distinct resonances: a higher-energy repulsive polaron (RP) and a lower-energy attractive polaron (AP). These features arise from the interaction between excitons and the Fermi sea and serve as robust optical signatures of the local doping conditions in each layer.

In particular, RP and AP resonances exhibit similar spectral shapes for electron and hole doping in MoSe\textsubscript{2}, making it difficult to identify the doping polarity based solely on line shape analysis. To address this, we infer the carrier type using a gate capacitor model: decreasing $V_\mu$ lowers the chemical potential, resulting in hole doping, while increasing $V_\mu$ raises the chemical potential, leading to electron doping.

An exception to this ambiguity arises in Device~3, which consists of WSe\textsubscript{2} and MoS\textsubscript{2} layers \cite{roch2020first,wang2020observation}. In this device, electron and hole doping in the MoS\textsubscript{2} layer produce distinctly different optical responses. Under electron doping, two attractive polaron resonances (AP) emerge, while only a single AP feature is observed in the hole-doped regime, allowing for differentiation between the two doping types (see Fig.~\ref{fig:data_linecuts7}c,d).

\begin{figure*}
\includegraphics{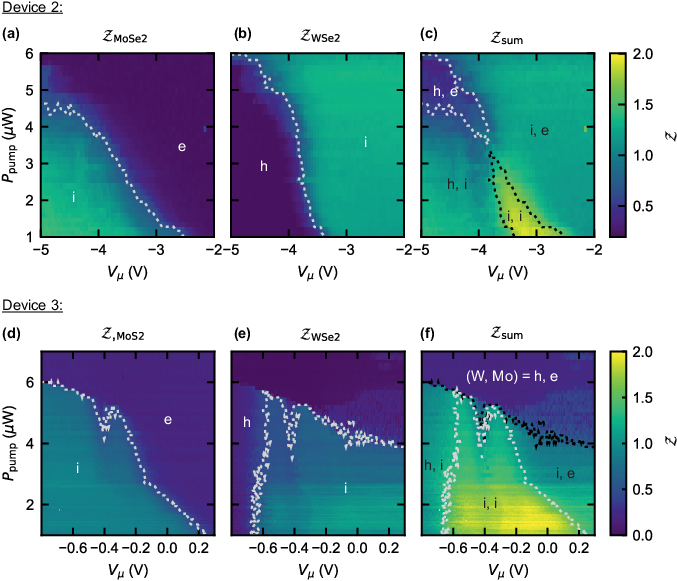}
\caption{\label{fig:data_differentdevice}
(a,b) Pump power ($P_\mathrm{pump}$) vs.\ gate voltage ($V_\mu$) maps of the peak 1s X (RP) reflection contrast of MoSe\textsubscript{2} (a) and WSe\textsubscript{2} (b) layers normalized to the individual reflectance contrast of each layer in the absence of charges ($\mathcal{Z}$) measured in Device 2. Regions corresponding to intrinsic (i), electron (e), or hole (h) doping are labeled. Black dashed contours indicate where $\mathcal{Z}$ falls below 0.5. (c) Overlay of the maps in (a) and (b) showing the full charge configuration in the heterostructure. Each region is labeled by the doping condition in the two layers. The black dotted contour marks the charge-neutral region (i,i), while the white dotted contour highlights regions of opposite doping (e,h or h,e), where both layers show enhanced $\mathcal{Z}$ above 0.5.
(d,e) Pump power ($P_\mathrm{pump}$) vs.\ gate voltage ($V_\mu$) maps of the peak 1s X (RP) reflection contrast of MoS\textsubscript{2} (d) and WSe\textsubscript{2} (e) layers normalized to the individual reflectance contrast of each layer in the absence of charges ($\mathcal{Z}$) measured in Device 3. Regions corresponding to intrinsic (i), electron (e), or hole (h) doping are labeled. Black dashed contours indicate where $\mathcal{Z}$ falls below 0.5. (f) Overlay of the maps in (d) and (e) showing the full charge configuration in the heterostructure. Each region is labeled by the doping condition in the two layers. The black dotted contour marks the charge-neutral region (i,i), while the white dotted contour highlights regions of opposite doping (e,h or h,e), where both layers show enhanced $\mathcal{Z}$ above 0.5.
}
\end{figure*}

Panels~\ref{fig:doping}c,d display two-dimensional maps of the peak 1s X (RP) reflection contrast of MoSe\textsubscript{2}/ WSe\textsubscript{2} layers normalized to the individual reflectance contrast of each layer in the absence of charges ($\mathcal{Z}$), plotted as a function of gate voltage $V_\mu$ and pump power $P_\mathrm{pump}$. At each point, $\mathcal{Z}$ was extracted from the resonance peak intensity, dynamically identified within a \qty{\pm 4}{\milli\electronvolt} window centered on the exciton energy under that condition. This procedure accounts for energy shifts due to repulsive polaron formation. The corresponding energy ranges are illustrated in Fig.~\ref{fig:data_linecuts7}a,b. Regions of intrinsic (i), electron (e), and hole (h) doping are labeled. Black dashed contours indicate where the residue falls below 0.7, and are used as a threshold to distinguish doped from undoped regimes.

In a next step, the two maps were then summed (Fig.~\ref{fig:doping}e), providing a comprehensive view of the charge configuration across the heterostructure:

\begin{equation}
\mathcal{Z}_{\mathrm{sum}}(V_\mu, P_\mathrm{pump}) = \frac{I^{\mathrm{Mo}}_{\mathrm{peak}}(V_\mu, P_\mathrm{pump})}{I^{\mathrm{Mo}}_{\mathrm{peak}}(\text{neutral})} + \frac{I^{\mathrm{W}}_{\mathrm{peak}}(V_\mu, P_\mathrm{pump})}{I^{\mathrm{W}}_{\mathrm{peak}}(\text{neutral})}.
\end{equation}

\begin{figure*}[!ht]
\includegraphics{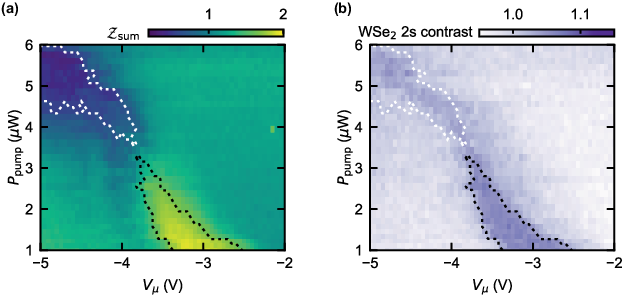}
\caption{\label{fig:data_linecuts9}
(a, b) $\mathcal{Z}_{\mathrm{sum}}$ (a) and WSe\textsubscript{2} 2s contrast map (b) shown as a function of gate voltage \(V_\mu\) and pump power \(P_\mathrm{pump}\). The data was measured on Device~2.
}
\end{figure*}

\begin{figure*}[!ht]
\includegraphics{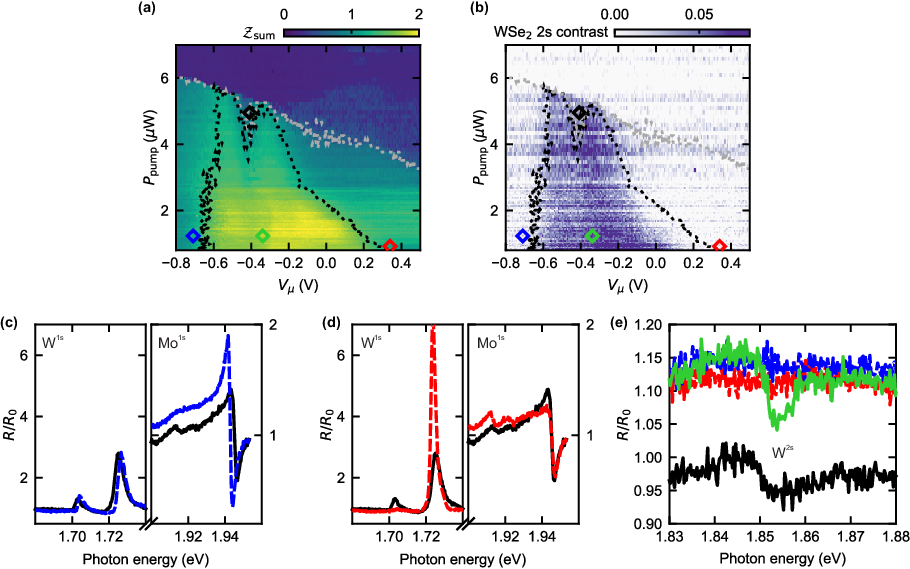}
\caption{\label{fig:data_linecuts7}
(a, b) $\mathcal{Z}_{\mathrm{sum}}$ (a) and WSe\textsubscript{2} 2s contrast map (b) shown as a function of gate voltage \(V_\mu\) and pump power \(P_\mathrm{pump}\). The data was measured on Device~3.
(c,d) Blue and red spectral linecuts show WSe\(_2\) in the hole‑doped regime (MoS\(_2\) is charge neutral) and MoS\(_2\)  in the electron‑doped regime (WSe\(_2\) is charge neutral), respectively, while the black lines show the spectrum in a regime where both layers are doped to the same extent. 
(e) $\mathrm{d} (R/R_0) / \mathrm{d} E $ in the Rydberg exciton regime, with the green line indicating the resonances at charge neutrality (see green diamond in (b)). The bare 2s resonances vanish when only one layer is doped (red and blue lines) but remain finite, without spectral shift, when both layers are doped at the same time (black line).
}
\end{figure*}

Each region is labeled according to the doping condition in the MoSe\textsubscript{2} and WSe\textsubscript{2} layers. The green dotted contour highlights the fully charge neutral region (i, i), where both layers exhibit reduced exciton residue below the 0.7 threshold. The black dotted contour delineates the region of opposite doping, where the individual exciton residue in both layers exceeds 0.7, indicating simultaneous electron and hole doping in the respective layers. In the following and throughout the main text, this summed version is referred to as the $\mathcal{Z}_{\mathrm{sum}}$ , denoting the combined exciton residues from both layers. Figure ~\ref{fig:data_differentdevice} presents two-dimensional maps of $\mathcal{Z}_{\mathrm{sum}}$ for Devices~2 and 3, respectively, analogous to the data shown for Device~1 in Fig.~\ref{fig:doping}. Analogous to the 1s $\mathcal{Z}_{\mathrm{sum}}$ map, WSe\textsubscript{2} 2s contrast maps are shown in Fig. 2 (Device~1), \ref{fig:data_linecuts9} (Device~2), and \ref{fig:data_linecuts7} (Device~3). The 2s contrast is defined as the peak-to-peak reflectivity difference of the dispersive Lorentzian lineshape associated with the WSe\textsubscript{2} 2s exciton.

\begin{figure*}[!ht]
\includegraphics{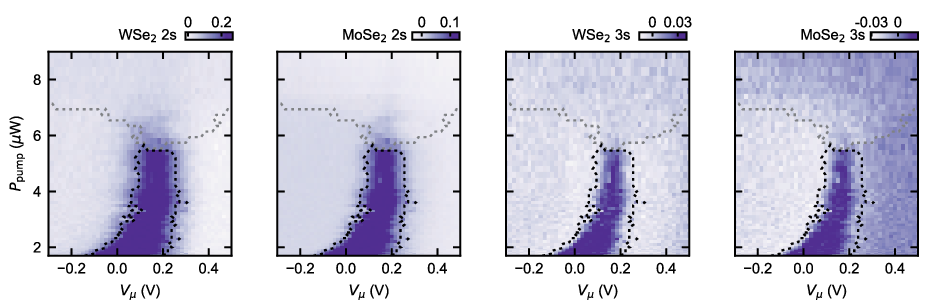}
\caption{\label{fig:data_Rydberg}
WSe\(_2\) 2s contrast (a), MoSe\(_2\) 2s contrast (b), WSe\(_2\) 3s contrast (c) and MoSe\(_2\) 3s contrast as a function of gate voltage \(V_\mu\) and pump power \(P_\mathrm{pump}\). The data was measured on Device~1. 
}
\end{figure*}

In all three devices, the 2s contrast signal extends beyond the region marked by the black dotted contour, remaining visible even where the 1s resonance becomes significantly weaker. Similar maps are also shown for the MoSe\textsubscript{2} 2s as well as the  WSe\textsubscript{2} and MoSe\textsubscript{2} 3s states (Fig. \ref{fig:data_Rydberg}). The 3s Rydberg exciton, with its larger Bohr radius compared to lower excitonic states \cite{liu2019magnetophotoluminescence}, undergoes the Mott transition at lower carrier densities, where interparticle spacing becomes comparable to the exciton size. This Mott density lies near the upper sensitivity limit of 1s polaron spectroscopy. As a result, by the time changes in the 3s contrast become visible, the carrier density is already too high to support a well-defined 3s resonance.

\section{Density Calibration}
\label{sec:carrier_density}

\begin{figure*}[ht!]
\includegraphics[scale=0.9]{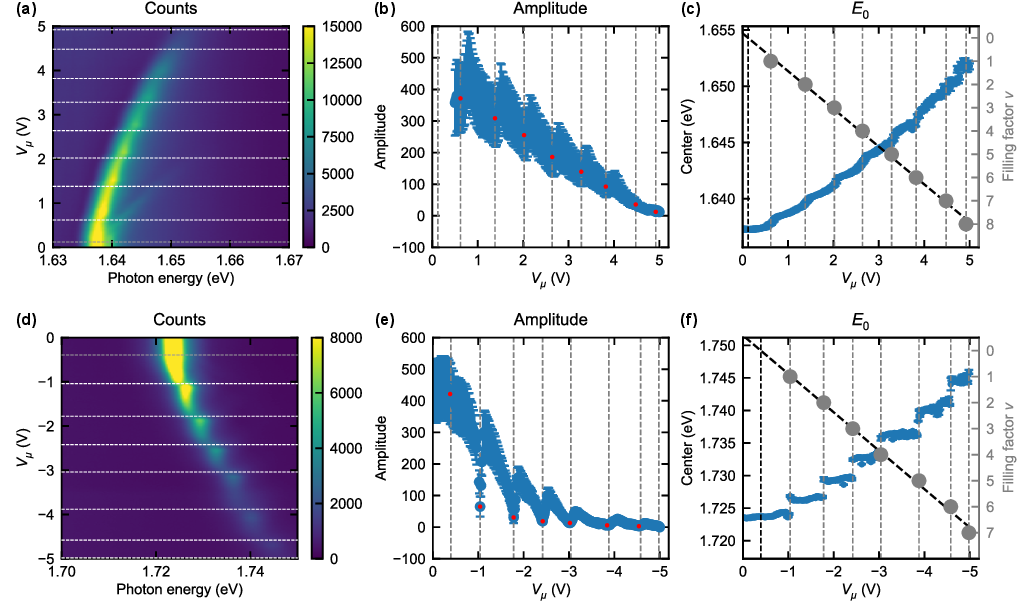}
\caption{\label{fig:ll_calibration}(a) Gate voltage \(V_\mu\) dependence of the reflection spectrum of the MoSe\(_2\) 1s exciton at $B = \qty{14}{\tesla}$ under $\sigma^+$ circular polarization. (b) Extracted amplitude and (c) center energy of the exciton resonance as functions of \(V_\mu\). White dashed lines in (a) and gray dashed lines in (b,c) indicate the positions of local minima in the amplitude (red markers), which coincide with abrupt shifts in the resonance energy. Gray markers denote gate voltages corresponding to integer carrier fillings. (d–f) Analogous measurements for the WSe\(_2\) 1s exciton at $B = \qty{14}{\tesla}$ under $\sigma^-$ circular polarization.
}
\end{figure*}

\begin{figure*}[ht!]
\includegraphics[scale=0.9]{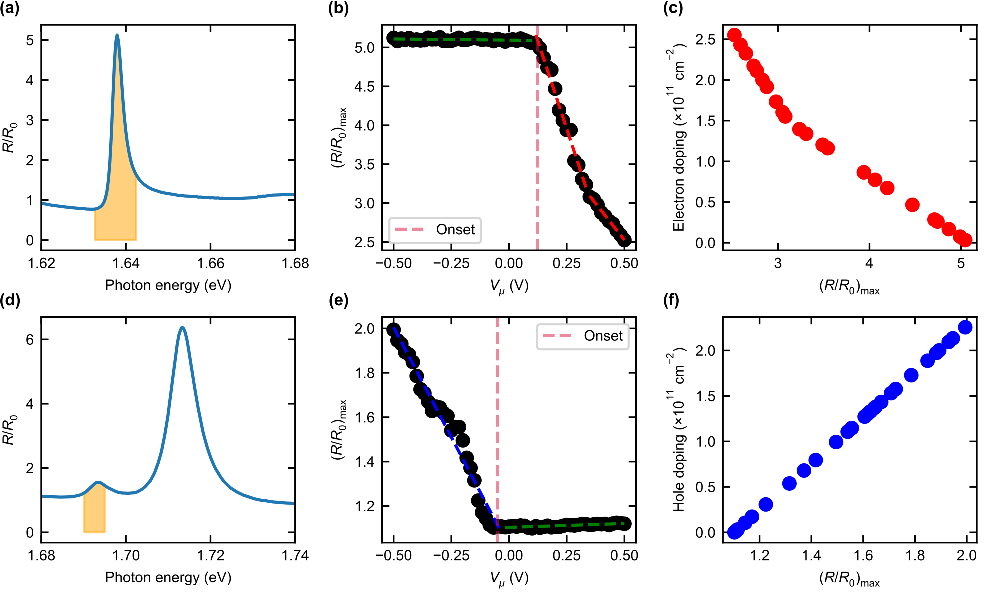}
\caption{\label{fig:optical_density_estimator}(a) Normalized reflection spectrum at gate voltage $V_\mu$= $\qty{-0.22}{\volt}$ near the MoSe\(_2\) 1s exciton.  (b) $(R/R_0)_{\mathrm{max}}$ as a function of $V_\mu$, showing the onset of electron doping in MoSe\(_2\). A phenomenological fit captures its gate dependence. is defined as the peak normalized reflection within the yellow-shaded energy window around the repulsive polaron resonance.
 (b) $(R/R_0)_{\mathrm{max}}$ as a function of $V_\mu$, showing the onset of electron doping in MoSe\(_2\) . A phenomenological fit captures its gate dependence.
 (c) Electron density $n_e$, calibrated via Landau level features (Fig.~\ref{fig:ll_calibration}c), plotted against $(R/R_0)_{\mathrm{max}}$.
 (d) Normalized reflection spectrum at $V_\mu$= $\qty{-0.22}{\volt}$ near the WSe\(_2\) 1s exciton. $(R/R_0)_{\mathrm{max}}$ is defined analogously as the peak normalized reflection within the yellow-shaded window around the attractive polaron resonance.
 (e) $(R/R_0)_{\mathrm{max}}$ as a function of $V_\mu$, showing the onset of hole doping in WSe\(_2\) , with a corresponding phenomenological fit.
 (f) Hole density $n_h$, calibrated via Landau level features (Fig.~\ref{fig:ll_calibration}f), plotted against $(R/R_0)_{\mathrm{max}}$.
}
\end{figure*}

In this section, we detail the procedure used to extract electron and hole densities ($n_e$ and $n_h$) from the optical response of the system across the full $V_\mu$–$P_{\text{pump}}$ parameter space. The goal is to provide a scheme that links optical observables, specifically the strength of the AP and RP resonances, to the underlying charge carrier densities.

Knowing the thicknesses of our hBN flakes, one can in principle estimate the charge carrier density using a parallel-plate capacitor model. In this approximation, the geometric capacitance per unit area is given by \( C_\mathrm{geom} = \varepsilon_0 \varepsilon_{\perp}^\mathrm{hBN} / t_{\perp} \), where \( t_{\perp} \) denotes the thickness of the (top or bottom) hBN layer, and \( \varepsilon_{\perp}^\mathrm{hBN} \) is its out-of-plane static dielectric constant.

To account for potential systematic uncertainties arising from sample-dependent variations in \( \varepsilon_{\perp}^\mathrm{hBN} \), we employed an alternative calibration method for Device~1. Specifically, to map the gate voltage \( V_\mu \) to carrier density in the absence of optical excitation, we utilized spectral signatures associated with integer quantum Hall states. These states manifest as discrete cusps and energy shifts in the 1s exciton resonance under a magnetic field of \( B = \qty{14}{\tesla} \) (Fig.~\ref{fig:ll_calibration}a,d), corresponding to changes in Landau level filling~\cite{smolenski2019interaction}. The density spacing between singly degenerate Landau levels is given by \( n = eB/h = \qty{3.39e11}{\per\square\centi\metre} \), where \( e \) is the elementary charge and \( h \) is Planck’s constant. Consequently, the voltage spacing between successive spectral features provides a direct calibration of the carrier density per unit \( V_\mu \). We systematically identify these transitions by locating local minima in the exciton amplitude, which coincide with sharp jumps or cusps in the resonance energy (Fig.~\ref{fig:ll_calibration}b,c,e,f).

\begin{figure*}[ht!]
\includegraphics{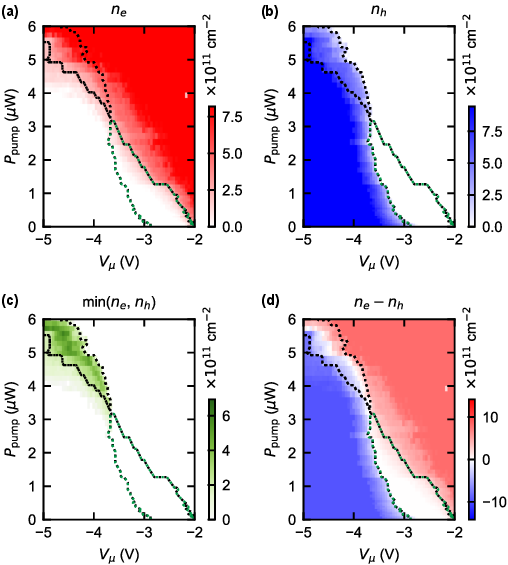}
\caption{\label{fig:stec9_density_calibration} Charge carrier densities extracted from the optical spectra measured in Device 2 as a function of the gate voltage, \(V_\mu\), and pump power, \(P_{\rm pump}\): (a) electron density, \(n_e\), in the MoSe\(_2\) layer; (b) hole density, \(n_h\), in the WSe\(_2\) layer; (c) minimum of \(n_e\) and \(n_h\); (d) excess charge \(n_e - n_h\). Dotted contour lines denote the same boundaries depicted in Fig. \ref{fig:data_linecuts9}.
}
\end{figure*}

Having established an electrical calibration of the gate voltage using Landau levels, we next developed a purely optical scheme to track the charge carrier density under non-resonant optical pumping. To this end, we analyzed the normalized reflection spectra around the 1s excitons of both MoSe$_2$ and WSe$_2$ for Device~1 (Fig.~\ref{fig:optical_density_estimator}). For each spectrum recorded at zero pump power, we identify the peak value within a narrow energy window around the polaron resonance and denote it as $(R/R_0)_{\mathrm{max}}$ (yellow shading in Fig.~\ref{fig:optical_density_estimator}a,d). Fig. \ref{fig:optical_density_estimator}b shows how $(R/R_0)_{\mathrm{max}}$ evolves with the electrostatic gate voltage $V_\mu$ on the MoSe\(_2\) side: a pronounced decrease marks the onset of electron doping (based on the drop in RP strength, as previously discussed in Ref.~\cite{kiper2025confined}.), which we capture with a simple phenomenological fit. By combining this optical observable with the Landau-level density calibration introduced above (Fig.~\ref{fig:ll_calibration}), we obtain a direct mapping between $(R/R_0)_{\mathrm{max}}$ and the electron density $n_e$, plotted in Fig. ~\ref{fig:optical_density_estimator}c. The procedure is repeated for WSe\(_2\), where $(R/R_0)_{\mathrm{max}}$ tracks the AP resonance (Fig.~\ref{fig:optical_density_estimator}d).  $V_\mu$ dependence (Fig.~\ref{fig:optical_density_estimator}e) similarly pinpoints the onset of hole injection, and the corresponding calibration against Landau levels yields the relation between $(R/R_0)_{\mathrm{max}}$ and the hole density $n_h$ (Fig.~\ref{fig:optical_density_estimator}f). Together, these calibrations provide a purely optical procedure for quantifying the charge carrier density across the full range of gate voltages and pump powers explored.

The charge carrier densities for Device~1 are presented in Fig.~3 of the main text, while Fig.~\ref{fig:stec9_density_calibration} shows the corresponding maps for Device~2. A similar procedure was followed for Device~2, with the only difference being that the initial gate voltage calibration was performed using a parallel-plate capacitor model. Due to its thicker hBN spacer, Device~2 allows access to higher carrier densities.

\section{Microscopic Origin of Asymmetric Doping}
\label{sec:carrier_density}

\textcolor{black}{MoSe$_2$–WSe$_2$ heterobilayers feature a type-II band alignment, with the conduction-band minimum in MoSe$_2$, the valence-band maximum in WSe$_2$, and an interlayer gap of approximately \(\qty{1.6}{\electronvolt}\) \cite{kang_band_2013, rivera2018interlayer}. This asymmetric doping behavior can be understood microscopically: density functional theory and ARPES studies show that non-resonant optical pumping generates electron-hole pairs in both layers and the dominant ensuing process is intralayer radiative recombination. However, holes generated in MoSe$_2$ have a finite probability to be transferred to WSe$_2$ via $\Gamma$-valley states \cite{wilson2017determination}; this process is expected to occur irrespective of the twist angle between MoSe$_2$ and WSe$_2$ and even across few-layer hBN spacer layers \cite{shimazaki_strongly_2020, yoon2022charge}. In contrast, electrons in MoSe$_2$ relax to the lowest energy K and K’ valleys in the same layer and form a long-lived bound interlayer exciton state with a hole in WSe$_2$. It is plausible that the electron–hole pair in MoSe$_2$ remains bound throughout the interlayer hole transfer process. Furthermore, the interlayer charge transfer mechanism we described naturally accounts for the preferential accumulation of holes in WSe$_2$ and for the bending of the $(i,i)$ boundary towards the hole doping side at low pump powers, where mid-gap defect states in MoSe$_2$ can further enhance hole generation through defect-assisted exciton dissociation \cite{handa2024spontaneous}.}

\section{Further comments on the 2s exciton in the presence of interlayer excitons}
\label{sec:theory}
\textcolor{black}{Consider first the interactions between excitons and electrons or holes. In second-order perturbation theory the leading-order interaction is a charge-induced dipole interaction: the charge polarizes the exciton and then interacts with the induced dipole. The strength of this interaction depends on the polarizability $\alpha$ of the exciton and asymptotically scales as
\begin{equation}
    V_{\alpha}(R) =  -\frac{\alpha}{2R^4} \  \mathrm{for} \ \ R \rightarrow \infty.
\end{equation}
The polarizability of an exciton in state $i$ is given by
\begin{equation}\label{eq:polarizability}
\alpha_i =2 \sum_{n\neq i}\frac{\mu_{ni}^2}{E_n-E_i},
\end{equation}
where $\mu_{ni}$ is the transition dipole moment. Since the 2p-states lie just lower in energy than the 2s state, the polarizability of the 2s-state is negative \cite{christianen2025hybridzation}. It therefore has repulsive interactions with charges, and it will behave as a repulsive polaron. Its attractive polaron branch arises from the hybridization with the 2p state.}

\begin{figure*}[ht!]
\includegraphics[width=0.75\textwidth]{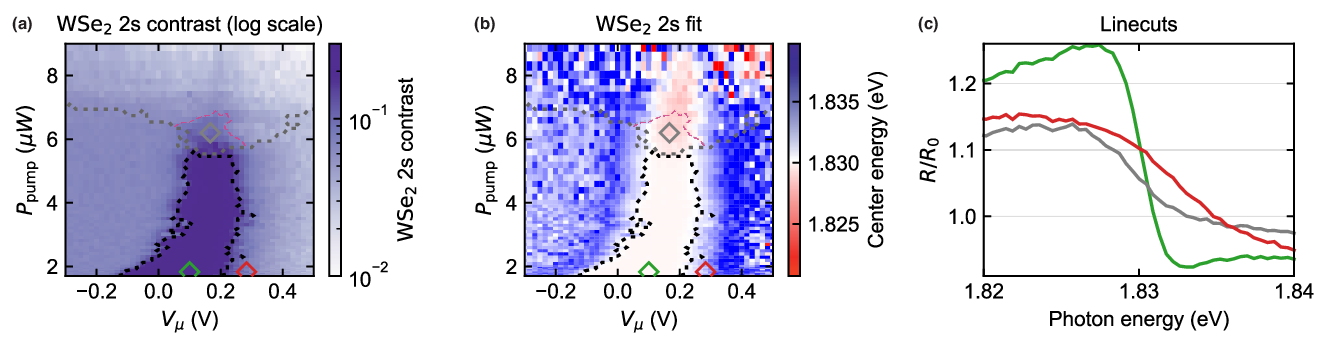}
\caption{\label{fig:2sfit} (a) Log-scale colormap of the WSe\textsubscript{2} 2s contrast as a function of $V_\mu$ and $P_\mathrm{pump}$ (same dataset as Fig.~2c).
The dotted grey and black contours indicate regions with distinct charge configurations, defined as in Fig.~1e.
The dashed pink contour marks the boundary where the WSe\textsubscript{2} 2s contrast drops below 0.055 ($20\%$) within the region of opposite doping between the layers.
(b) Colormap of the fitted center energy of the WSe\textsubscript{2} 2s resonance, obtained from dispersive Lorentzian fits. The 2s exciton shows a small red-shift in the presence of interlayer excitons.
(c) Representative reflection spectra ($R/R_0$) across the 2s WSe\textsubscript{2} Rydberg exciton resonance.
The green curve (measured near charge neutrality, green diamond) shows the Rydberg exciton; the red curve, taken when the MoSe\textsubscript{2} layer is electron-doped, exhibits a blue-shift and broadening; the grey curve, measured under interlayer exciton formation, shows a red-shift and broadening of the 2s resonance.}
\end{figure*}

\textcolor{black}{Since interlayer excitons are neutral objects, their leading-order interaction with the 2s-exciton is of higher order. Furthermore, the out-of-plane dipole moment of the interlayer exciton cannot directly induce a dipole moment or hybridize the s- and p-states in the intralayer exciton. As a result, the most important interaction is the van der Waals interaction arising from correlated dipole fluctuations, which asymptotically scales as
\begin{equation}
V_{vdw}(R) = -\frac{C_6}{R^6} \ \mathrm{for} \ \ R \rightarrow \infty.
\end{equation}
The $C_6$ in this case is given by
\begin{equation}
    C_6=\frac{5}{4}\sum_{n_1, n_2 \neq i_1,i_2} \frac{\mu_{ni1}^2 \mu_{ni2}^2}{E_{n_1}+E_{n_2}-E_{i_1}-E_{i_2}}.
\end{equation}
In this model, the leading-order contribution to the interaction is the correlated excitation of both the intralayer and interlayer excitons to their respective 2p states. In this case $E_{2p,intra}+E_{2p,inter}-E_{2s,intra}-E_{1s,inter}>0$, resulting in an attractive interaction. Since the bound state energy of a possible trion is likely to be small, this means that the 2s-state forms a well-defined attractive polaron state with a small lineshift. Experimentally, we indeed observe that the 2s exciton exhibits a small redshift in the presence of interlayer excitons (see Fig.~\ref{fig:2sfit}b,c), while a clear blueshift and repulsive polaron is broadening in the presence of charges in either layer.}

\textcolor{black}{Another exciton line-broadening mechanism, aside from polaron broadening, originates from inelastic processes. Indeed, when the interlayer exciton and 2s exciton collide, the 2s exciton can decay into a 1s exciton, leading to the dissociation of the interlayer exciton. The 2s exciton can also decay into the 1s exciton due to interactions with electrons. At present, it is unclear how strong the broadening due to these inelastic decay processes is compared to the polaron broadening. However, these inelastic processes might be the reason the 2s exciton disappears for larger densities of interlayer excitons, even when the polaron broadening is potentially small for this line.}

\newpage

%\pagebreak

%\bibliography{Citations}% Produces the bibliography via BibTeX.

%